\documentclass[journal]{IEEEtran}
\usepackage{amsmath,amsfonts}
\usepackage{algorithmic}
\usepackage{algorithm}
\usepackage{array}
\usepackage[caption=true,font=normalsize,labelfont=sf,textfont=sf]{subfig}
\usepackage{textcomp}
\usepackage{stfloats}
\usepackage{url}
\usepackage{verbatim}
\usepackage{graphicx}
\usepackage{cite}

\usepackage{titlesec}
\titlespacing*{\subsubsection}{0pt}{0.5ex}{0.25ex}
\titlespacing*{\subsection}{0pt}{0.5ex}{0.25ex}
\titlespacing*{\section}{0pt}{0.5ex}{0.25ex}
\begin{document}

\title{{\fontsize{17pt}{16pt}\selectfont L--Moment–Based LOS/NLOS Channel Characterization via Four-parameter Kappa Distribution for AoA BLE CTE Measurements}}


\author{Hamed Talebian,
        Aamir Mahmood,~\IEEEmembership{Senior~Member,~IEEE}
        Mikael Gidlund,~\IEEEmembership{Fellow~Member,~IEEE}


\thanks{H. Talebian, A. Mahmood, and M. Gidlund are with the Department of Computer and Electrical Engineering, Mid Sweden University, 851 70, Sundsvall, Sweden (email: hamed.talebian@miun.se)}
\vspace{-20pt}
}


\maketitle

\begin{abstract}
Bluetooth Low Energy (BLE) CTE transmissions provide in-phase and quadrature (IQ) samples whose empirical statistics are strongly governed by the propagation regime; in particular, the distributions differ markedly between line-of-sight (LOS) and non-line-of-sight (NLOS) conditions. In NLOS, multipath-induced distortions typically degrade Angle-of-Arrivial (AoA) estimation accuracy. Existing BLE direction finding datasets rarely provide tightly controlled, IQ-level paired LOS/NLOS measurements with rigorous statistical validation, and commonly used flat-fading models can be inadequate for cluttered indoor environments exhibiting heavy-tailed power distributions. To address these limitations, we conduct a paired-geometry BLE AoA measurement campaign using an off-the-shelf module, collecting 132,000 labeled CTE packets under matched anchor–tag conditions. A robust preprocessing stage removes anomalous CTEs using combined univariate and multivariate criteria. Feature-wise hypothesis tests on IQ-derived power features confirm strong LOS/NLOS separability. All mean differences are statistically significant at $\mathbf{\mathrm{\alpha}} \le \mathbf{0.01}$, and adjusted $\mathbf{P} \le \mathbf{10^{-6}}$; additionally 92\% of feature-wise variance differences are significant. We further compute L--moment ratios (LMRs) and analyze them in the L--moment Ratio Diagram (LMRD), showing that NLOS subsets exhibit markedly heavier tails and stronger asymmetry than LOS. Kappa-family distributions fitted from LMRs provide substantially improved dual scored L--moment goodness-of-fit (GoF), Specifically, for NLOS with $\mathbf{D_{\mathrm{NLOS}}}=\mathbf{3.20\times 10^{-4}}$, which is the smallest discrepancy in the LMRD and a near-zero standardized L--kurtosis deviation ($\mathbf{Z_{\mathrm{NLOS}}}=\mathbf{1.18\times 10^{-2}}$). As a practice, we apply a self-supervised clustering to L--moment statistics, achieving a more separable representation, compared to product moments. To the best of our knowledge, this is the first application of L--moment analysis to BLE AoA CTE measurements, yielding compact, robust descriptors for AoA preprocessing, LOS/NLOS detection, and realistic simulation under heavy-tailed propagation. 
\end{abstract}


\section{Introduction}
\IEEEPARstart{B}luetooth Low Energy (BLE) direction-finding platforms have enabled large-scale deployment of lightweight sensing in indoor industrial, healthcare, and logistics environments, leveraging ubiquitous commercial hardware and low-power operation. In such systems, the in-phase and quadrature (IQ) samples, acquired during the Constant Tone Extension (CTE), are strongly governed by the propagation regime. Under line-of-sight (LOS), a dominant direct component typically yields more stable signal statistics, whereas non-line-of-sight (NLOS) conditions introduce blockage, scattering, and rich multipath that alter the IQ distributions through shadowing and phase-dependent superposition. These channel-induced effects appear directly at the IQ level and can severely impact downstream processing; therefore, reliable LOS/NLOS detection from CTE IQ measurements is a key enabling function for robust operation. As a practical ML-assisted solution, LOS/NLOS classification can be integrated as an auxiliary inference stage to condition preprocessing and modeling. Accordingly, AI-assisted estimation of propagation-state parameters—most notably LOS/NLOS classification—provides a mechanism for propagation-conditioned preprocessing and realistic channel modeling/simulation, improving robustness in cluttered indoor environments \cite{pan2025ai}.


Several BLE AoA datasets and tools have appeared in recent years. MATLAB tutorial and tools such as~\cite{MathWorksBLENodeTracking}, mainly focus on canonical channel impairments (AWGN, Carrier Phase Offset (CFO) compensation, etc.) and their impact on the AoA processing chain through simulation with synthetic IQ data and basic processing blocks. Indoor measurement campaigns with commercial receivers (e.g., \cite{Leitch2024BLEIQ, Maus2021AoAIndoorLoc, Paulino2022UCABLE}) report the AoA estimation performance under practical deployment scenarios, with a particular focus on localization error and phase-based AoA estimation. Complementary simulation frameworks \cite{Piazzese2021NURA} generate synthetic IQ data under configurable array geometries and channel models, with the goal of analyzing the Cramér–Rao bound. These contributions are valuable for evaluating AoA estimation algorithms and end-to-end localization performance, and they have advanced the experimental efforts in BLE direction finding.

However, from the perspective of IQ-level LOS/NLOS channel modeling, existing BLE AoA datasets and models share several limitations. \textit{Firstly}, LOS and NLOS measurements are typically not paired under strictly matched anchor–tag geometry; obstacles and blockages are incidental to the scenario rather than systematically configured to produce class-labeled LOS/NLOS realizations of the same spatial configuration. As a result, it is difficult to separate the impact of propagation conditions from changes in geometry. \textit{Secondly}, most analyses and simulation frameworks rely on additive white Gaussian noise or simple flat-fading assumptions, such as Rayleigh or Rician models, which may be inadequate in cluttered indoor environments where the amplitude statistics of IQ-derived power can be skewed and heavy-tailed. \textit{Thirdly}, although LOS/NLOS separation is often assumed in system-level studies, there is a lack of rigorously statistically validated LOS/NLOS labels at the IQ level for BLE AoA, especially in large-scale datasets suitable for machine learning (ML).

In parallel, there is increasing evidence from wireless channel measurements and other domains that heavy tails and outliers can severely impact classical product-moment-based statistics. In indoor and industrial radio channels, sporadic strong multipath components and deep fades result in empirical amplitude distributions with significant skewness and kurtosis that are poorly captured by light-tailed Gaussian models. In such regimes, ordinary higher-order moments can have large sampling variance, and even the existence of finite moments may be questionable. L--moments \cite{hosking1990moments} provide a robust alternative since they are linear functions of order statistics, less sensitive to extreme values, and yield dimensionless ratios such as L--skewness and L--kurtosis that summarize shape and tail behavior more reliably than conventional moments. When plotted in the L--skewness–L--kurtosis diagram/L--moment ratio diagram (LMRD), empirical L--moment ratios (LMRs) trace clusters that can be matched to flexible parametric families, such as the four-parameter Kappa distribution \cite{hosking1997regional}, which encompasses several generalized extreme-value and heavy-tailed models as special cases. While L--moments have been successfully applied in hydrology and network traffic modeling (e.g., \cite{galeano, aparicio, dey, torres2025applying}), their use for physical-layer characterization of a radio channel, and in particular for LOS/NLOS channel separation of BLE AoA CTE measurements has, to the best of our knowledge, not been explored.

These observations motivate the following problems. To design robust BLE AoA systems and data-driven LOS/NLOS detectors, we require: (i) a dataset of tightly paired LOS/NLOS IQ captures under identical geometry, collected with realistic BLE AoA hardware; (ii) a statistical pipeline that can rigorously validate LOS/NLOS separability at the IQ-derived feature level; and (iii) an IQ-level LOS/NLOS channel characterization that can capture the heavy-tailed and asymmetric behavior of indoor BLE power statistics beyond classical Rayleigh/Rice assumptions. 


In this work, we address these gaps by combining a controlled BLE AoA measurement campaign with an L--moment-based statistical modeling framework. Using an off-the-shelf BLE AoA platform in a realistic indoor environment, we provide a collection of CTE packets, containing IQ samples under strictly paired LOS/NLOS conditions for identical anchor–tag positions and orientations. A robust preprocessing pipeline removes anomalous CTE packets using a combined univariate and multivariate outlier detection approach. Afterward, feature-wise statistical hypothesis tests for the equality of the first and second moments of LOS and NLOS subsets are employed to validate LOS/NLOS separability across the full 64-dimensional IQ-derived power feature space. Building on this validated dataset, we estimate LMRs for each feature, analyze LOS and NLOS clusters in the LMRD, and fit the four-parameter Kappa family of distributions in the LMR domain. The fitted Kappa models are then compared with standard fading distributions in terms of L--moment-based goodness-of-fit (GoF), with particular emphasis on the heavy-tailed NLOS regime and its implications for AoA preprocessing and ML-based LOS/NLOS classification. In this respect, our main contributions are:

\begin{itemize}
    \item \textit{Paired BLE IQ dataset for LOS/NLOS channel classification:} We design and carry out a BLE AoA CTE measurement campaign in which LOS and NLOS captures are recorded under strictly identical anchor–tag positions and orientations, yielding a large-scale, class-labeled IQ dataset suitable for ML and channel characterization.
    \item \textit{IQ-level statistical validation of LOS/NLOS separation:} We develop robust outlier removal rules and apply feature-wise Welch and Levene tests to 64 IQ-derived power features, demonstrating strong and widespread LOS/NLOS differences in both mean and, for most features, variance across the antenna–sample grid.
    \item \textit{L--moment channel characterization and Kappa-family modeling}: We estimate the LMRs for each feature, demonstrating clear LOS/NLOS clusters in the LMRD, and show that Kappa-family probability distributions provide substantially better L--moment-based GoF, compared with traditional Rayleigh/Rice fading distributions, particularly in NLOS channels where the heavy-tail behavior dominates.
    \item \textit{Implications for ML-assisted AoA estimation and ML preprocessing:} We argue that the resulting L-moment/Kappa descriptors form compact, robust features for LOS/NLOS detection and realistic BLE AoA channel simulation, and can be integrated into AoA preprocessing chains and ML models to improve robustness under heavy-tailed channel conditions. As an illustrative example, we compare clustering performance using the conventional product moments versus the L--moment ratio within a DBSCAN framework. This comparison highlights the superior clustering capacity and stability of L--moment–based representations, demonstrating the effectiveness of LMRs analysis for downstream ML tasks.

\end{itemize}

The structure of the paper is as follows. Section~\ref{sec:st} introduces the statistical framework, including IQ-power feature definition, outlier detection methods, hypothesis tests on first-order moments, and the L--moment/Kappa-based modeling tools. Section~\ref{sec:exf} details the experimental setup, hardware platform, indoor measurement environment, and the measurement campaign to generate paired LOS/NLOS CTE data. Section~\ref{sec:ExpResults} presents the experimental results, including exemplar analysis, hypothesis-testing outcomes, and LMRDs with fitted Kappa distributions. Finally, the concluding section summarizes the main findings and outlines future research directions.

\section{statistical framework}\label{sec:st}

The primary criterion for selecting statistical tools in this work is their ability to verify whether the experimental setup reliably captures two distinct dataset classes, namely LOS and NLOS. Because classical hypothesis-testing procedures are sensitive to outliers, we first introduce three complementary outlier-detection methods to ensure data integrity. We then review appropriate hypothesis-testing techniques to assess differences in first-order moments between the two classes. Finally, L--moment analysis is employed to validate the empirical distributional characteristics of the LOS and NLOS data subset and to identify the best-fitting distribution models.

In the context of direction-finding with Bluetooth 5.1, IQ samples refer to the in-phase (I) and quadrature (Q) components of the BLE received baseband signal (See. \cite{bleSignal}), captured over time by the receiver during the CTE processing, which is a new method to request and send short continuous wave (CW) signals within a normal Bluetooth packet after protocol data units (PDUs) field \cite{bluetooth}. It encompasses a constantly modulated series of unwhitenened 1's, each of which contains the IQ samples. These samples preserve both amplitude and phase information of the incoming RF wave, thereby enabling computation of relative phase differences across antenna elements for BLE AoA localization. Given the antenna layout and switching pattern, a series of non-repeated IQ samples in a CTE measurement is generated. The absolute value of each IQ sampling is considered a feature in this manuscript, and a scalar power feature is defined by  
\begin{equation}
  X_{j}^{(n)} = \bigl| I_{j}^{(n)} + jQ_{j}^{(n)} \bigr|^2
      = \bigl|(I_{j}^{n})^2 + j(Q_j^{n})^2\bigr|
      \label{eq:IQ_sample}.
\end{equation}
Mapping $10\mathrm{log_{10}(.)}$ of \eqref{eq:IQ_sample} gives the power figures in dB scale. Collecting a $k$-dimensional column vector gives the features as $\mathbf{X}^{(n)}\;=\;(\mathbf{X}_1^{(n)}, \cdots, \mathbf{X}_{k}^{(n)})$. 
For each IQ sampling index $j \in \{1, \cdots, k\}$, the set $\{X_j^{(n)}\}_{n =1}^{N}$ collects $\mathrm{N}$ CTE packets for LOS and NLOS measurements to form univariate observations for hypothesis testing and L--moment estimation at that antenna-sample index pairs.

\subsection{Outlier Detectors}
\label{subsec:outlier_detector}

Three outlier detectors are employed because they are sensitive to different types of anomalies in the CTE IQ power data. The inter-quantile range (IQR) rule is a simple univariate filter that flags observations that lie far outside the inter-quartile fences in each feature independently, but it can be sensitive to skewness and heavy tails. As shown in \cite[~p.56]{dodge2008concise}, $X_j^{(n)}$ is an outlier if $X_j \notin [\,q_1 - k\,\mathrm{IQR},\, q_3 + k\,\mathrm{IQR}\,]$ and $\mathrm{IQR}=q_3-q_1$, and the $k$ coefficient is set to 1.5 for mild outlier detection. This rule captures incidental anomalies introduced during the measurement campaign (e.g., sudden temporal changes in the propagation environment due to moving obstacles, transient interference from co-located short-range wireless devices, short-lived hardware or synchronization glitches, etc.), which do not necessarily reflect systematic or physically meaningful differences between the LOS and NLOS subset datasets.

Additionally, the Median Absolute Deviation (MAD)-based method is a univariate yet more robust and distribution-free scale estimator, as it uses the median and a scale-invariant deviation measure, allowing it to detect extreme deviations even when the marginal distributions are non-Gaussian. MAD is defined by 
\begin{equation}
\begin{aligned}
\mathrm{MAD}_{j} &=\operatorname{median}\bigl(|X_{j}^{(n)}-\tilde{X_j}|\bigr)\\[3pt]
z_j^{(n)} &=\frac{\alpha\left(X_{j}^{(n)}-\tilde{X_j})\right)}{\mathrm{MAD_j}}.
\end{aligned}
\label{eq:MAD}
\end{equation}
In Eq.~\eqref{eq:MAD}, $\tilde{X_j} = \operatorname{median}(\{X_j^{(n)}\}_{n=1}^N)$, the scale factor of $\alpha$ and threshold ($\tau$) setting distinguish how outliers are flagged. 

In contrast, the Minimum Covariance Determinant (MCD)-based outlier detection is explicitly multivariate. It measures how far each $k$-dimensional observation (column) vector lies from a robustly estimated joint center given the covariance structure, thus, it identifies points that are anomalous in their combination of feature values even if no single feature is individually extreme. It minimizes the weights of an appropriate function to approximate the means and the covariance matrix, then computes the Mahalanobis distance \cite{hubert2018minimum} as defined by 
\begin{equation}   
d_j^2=(\mathbf{X}^{(n)}-\hat{\mu}_{\mathrm{MCD}})^\top \hat{\Sigma}_{\mathrm{MCD}}^{-1}(\mathbf{X}^{(n)}-\hat{\mu}_{\mathrm{MCD}}).
\label{eq:mahalanobis}
\end{equation}
MCD statistic appropriator selects an h-subset size (based on dataset statistics) for minimization, which is a high breakdown point while maintaining a favorable ratio between the number of packets and IQ features. Therefore, the resulting robust statistic estimates and Mahalanobis distances can be interpreted as stable summaries of the joint feature space. Under the elliptically symmetric assumption, an observation is an outlier if its $d_j^2$ is greater than a chi-square distribution with $p$ degrees of freedom and an acceptable quantile.

The IQR rule primarily flags outliers that arise from random fluctuations in individual physical measurements, whereas the MAD rule is more effective at identifying observations that contribute to heavy tails in the marginal (per-feature) distributions. In addition, the MCD is well-suited when treating each LOS and NLOS label as realizations from a joint empirical distribution, as it captures multivariate deviations from the robust center and covariance structure. Consequently, when all three rules are applied in combination, the remaining observations predominantly represent the central statistical tendency of each subset, yielding cleaner LOS and NLOS datasets that are more suitable for subsequent analysis, albeit at the cost of a substantial reduction in CTE measurement size.

\subsection{Statistical Tests for Ordinary Moments}
\label{subsec:statistical_tests}
For each column feature vector $\mathbf{X_j^{(n)}}$, the observations are separated into class-conditional with sample means $\bar{X}_{j}^{m}$ and variances $(S_{j}^{m})^2$ and $m \in M=\{0,1\}$, corresponding to LOS and NLOS subset. To test equality of class-wise mean difference, Welch’s unequal variance $t$–statistic ($t_j$) and degrees of freedom (DoF) ($\nu_j)$ \cite[sec. 10.2]{devore2012modern} is given by 
\begin{equation}
t_j=\frac{\bar{X}_j^{1}-\bar{X}_j^{0}}{\sqrt{\tfrac{(S_j^{1})^2}{N_j^{1}}+\tfrac{(S_j^{0})^2}{N_j^0}}}\quad
\nu_j=\frac{\bigl(\tfrac{(S_j^1)^2}{N_j^1}+\tfrac{(S_j^0)^2}{N_j^0}\bigr)^2}{\tfrac{(\tfrac{(S_j^1)2}{N_j^1})^2}{N_j^1-1}+\tfrac{(\tfrac{(S_j^0)^2}{N_j^0})^2}{N_j^0-1}}.
\label{eq:tTest}
\end{equation}
Under null hypothesis, $H_0:\bar{X_j}^{0}=\bar{X_j}^{1}$, the two-sided $p$–values is compared with the t-distribution. As a result, K hypothesis tests are applied simultaneously to reject or accept of mean equivalency. If rejected, the mean value difference between LOS and NLOS subsets is statistically significant, and it can be confirmed that they are drawn from different distributions. Similarly, Levene’s test \cite{zhou2023statistical} is applied on the absolute IQ values to assess equality of variances, by forming absolute deviations, as $Z_{j}^{n} = \bigl|X_{j}^{(n,m)}-T_j^{m}\bigr|$ where $T_j^{m}$ is the group center (e.g., mean) for each class in $M$, and then the Analysis of Variance (ANOVA)-like statistic is computed by 
\begin{equation}
W=\frac{(N-2)N_j^{(n),0}(\bar{Z_j}^{(n),0}) + N_j^{(n),1}(\bar{Z_j}^{(n),1})}
{\sum_{j=1}^{N_j}\bigl(Z_{j}^{(n),0}-\bar{Z_{j}}^{(n),0}\bigr)^2+ \sum_{j=1}^{N_j}\bigl(Z_{j}^{(n),1}-\bar{Z_j}^{(n),1}\bigr)^2}. 
\label{eq:Wstats}
\end{equation}
In Eq.~\eqref{eq:Wstats}, $\bar{Z}_j^{(n),m}$ is the overall mean of the $Z_{j}$, and $N_j$ is the number of observations on each group (LOS/NLOS). The null hypothesis of equal variances, $H_0:(\sigma_j^0)^2=(\sigma_j^1)^2$, is compared with F-distribution and rejected at $\alpha$ level if $W>F_{1-\alpha;N-2}$ (equivalently, associated $p\le\alpha$).
Additionally, Benjamini–Hochberg false discovery rate (BH-FDR) procedure \cite{benjamini1995controlling} is used as a correction metric for parallel hypothesis testing that controls the expected proportion of false rejections among all rejected hypotheses. Given $m$ $p$-values, sorted as $p_{(1)}\le\cdots\le p_{(m)}$, BH-FDR rejects all hypotheses up to the largest index $k$ satisfying $p_{(k)}\le \frac{k}{m}q$, where $q$ is the target BH-FDR level. First moment inequality tests rely on the asymptotic normality of observations. In our case, each IQ-derived feature statistic is estimated from several thousand CTE packets per class; thus, the central limit theorem guarantees that the test statistics are well approximated by asymptotic distributions even when the underlying marginal power distributions are non-Gaussian. Together, these tests indicate for which features both the central tendency and dispersion of LOS and NLOS samples are statistically significant, thereby confirming that the two classes correspond to different underlying propagation conditions and justifying the use of class-specific modeling and classification.

\subsection{L--Moments Statistics}
L--moment analysis is a framework for describing, estimating, and testing probability distributions based on L--moments, which are expectations of linear combinations of order statistics rather than integer powers of first-order statistics. Formally, the (r)-th L--moment ($\lambda_r$) exists for any random variable (r.v) with finite mean and uniquely characterizes the underlying distribution \cite{hosking1997regional}. Because L--moments are linear in the data, they are substantially less sensitive than the conventional moments to sampling variability and outliers, and they remain well-behaved for heavy-tailed distributions where higher ordinary moments may not exist or be poorly estimated. This robustness makes L--moment analysis particularly important for modeling extremes and small samples, enabling more reliable parameter estimation, quantile estimation, and GoF.

\subsubsection{L--moment Definition} \label{lmomentDerivatives}
Let $X$ be a real-valued random variable (r.v.) with cumulative distribution function (CDF) of $F(x) = P_{r}[X \le x]=u$ and (left-continuous) quantile function as defined by $Q(u)=\inf\{x:u \le F(x)\}$ and $u\in(0,1)$. The special case for a strictly increasing CDF is $Q(u)=F^{-1}(x)$ and $F(Q(u))=u$. The (r)-th L--moment, denoted by $\lambda_r$, represents the expected value of a specific linear combination of sample order statistics as defined by
\begin{equation}
\begin{aligned}
\lambda_r = \frac{1}{r}\sum_{k=0}^{r-1}(-1)^k\binom{r-1}{k}\,\mathbb{E}[X_{r-k:r}] \\
=\int_0^1 Q(u)\,P_{r-1}^*(u)\,du\quad r \in \mathbb{Z}_{\ge 1}.
\end{aligned}
\label{eq:lmr-definition}
\end{equation}
Equivalently in Eq.\eqref{eq:lmr-definition}, $\lambda_r$ can be expressed in the quantile-integral form, where $Q(u)$ denotes the quantile function of the underlying distribution, and $P_{r}^*(u)=P_{r}(2v-1)$ is the 
(r)-th normalized and shifted Legendre polynomial of degree $r$ on the unit interval \cite{hosking1990moments}. The transformation $u=2v-1$ rescales the standard Legendre polynomial $P_{r}(u)$ originally orthogonal on $[-1,1]$ to the probability domain $u\in(0,1)$, thereby preserving its orthogonality and enabling direct application to L--moment theory. It can be shown that Eq.~\eqref{eq:lmr-definition} has a discrete form defined by
\begin{equation}
\lambda_{r+1}=\sum_{k=0}^{r}p_{r,k}^*\alpha_k = \sum_{k=0}^{r}p_{r,k}^*\beta_k.
    \label{eq:AlphaBeta}
\end{equation}
In Eq.~\eqref{eq:AlphaBeta}, the polynomials of $\alpha_k \;= \;\int_0^1\; Q(u)(1-u)^{r}du$, and $\beta_k\;=\;\int_0^1\; Q(u)u^{r}du$, are probability weighted moment of a r.v., interpreted as an integrals of quantile function of $Q(u)$, weighted by the equivalent polynomials \cite{hosking1997regional}, $p_{r,k}^*$, defined by
\begin{equation}    
P_{r}^*(u)\;=\; \;\sum_{k=0}^{r}p_{r,k}^*u^k\;=\;\;\sum_{k=0}^{r}
\frac{(-1)^{r-k}(r+k)!}{(k!)^2\,(r-k)!}\;u^{k}.
\label{eq:LegendrePolinomials}
\end{equation}
The unique orthogonal shifted Legendre polynomials used in Eq.~\eqref{eq:LegendrePolinomials}, has an explicit form, which provides closed-form coefficients for all L--moment kernels and facilitates analytical evaluation of $\lambda_r$ when the quantile function $Q(u)$ is known. This explicit basis ensures that L--moments preserve orthogonality, linearity, and numerical stability across a broad range of distributions, including those with heavy tails. The first four L--moments of $\lambda_{r} \in \{1,2,3,4\}$ play roles analogous to location, scale, skewness, and kurtosis, respectively. A single extreme observation can only affect $\hat{\lambda}_r$ through a bounded weight, in contrast with conventional higher order moments, where an outlier with a large value is raised to a high power and can dominate the estimate. As a result, L--moments have smaller sampling variance and bias under heavy tails and in the presence of contamination. 

\subsubsection{L--moment Ratios} \label{lmomentratios}
To obtain scale-free measures of distributional shape, L--moment ratios (LMRs) are defined as
\begin{equation}
\tau_2 = \frac{\lambda_2}{\lambda_1} \qquad 
\tau_3 = \frac{\lambda_3}{\lambda_2} \qquad 
\tau_4 = \frac{\lambda_4}{\lambda_2}.
\label{eq:Lmomentratios}
\end{equation}
These ratios generalize the concepts of coefficient of variation (CV), skewness, and kurtosis using linear—rather than polynomial—functionals of the distribution, yielding estimators with substantially improved robustness to outliers and heavy tails. The dimensionless ratios $(\tau_3,\tau_4)$ provide stable, scale–free measures of skewness and kurtosis over a much wider class of distributions than conventional moment ratios. The combination of existence under mild conditions and reduced sensitivity to extremes is the key reason why L--moment analysis is particularly effective for modeling heavy-tailed and extreme–value phenomena.

The theoretical L--moments are defined for analyzing common theoretical probability distributions, but they have to be estimated in practice from a finite number of observations/measurements. Given a sample $x_{1:n}$ with order statistics $x_{(1)}\le\cdots\le x_{(n)}$, the standard Probability Weighted Moment (PWM) estimator (See. Appendix~\ref{sec:appen-A}) is defined by \eqref{eq:beta_hat}. Then, L--moments and ratios are similarly generated by \eqref{eq:lmr-definition}, \eqref{eq:AlphaBeta},  and \eqref{eq:lambdafour}, with a change in variable name from $\lambda_r$ to $\ell_r$ for clarity (See Appendix~\ref{sec:appen:ratios}). $\ell_r$ is unbiased estimator of $\lambda_r$ for moderate to large sample size as mentioned in \cite{hosking1997regional} for $n \ge 1000$. The CTE measurements are sufficient to make sampling variability negligible for LOS/NLOS separation in the LMRD, and L--moments and LMRs estimators converge rapidly for heavy-tailed distributions.

\subsubsection{Kappa-family Empirical Distribution} \label{kappa}

Given the mildly heterogeneous nature of the feature distribution of LOS and NLOS subsets, identifying a single exact empirical parent distribution is not our primary objective. Instead, the goal is to obtain a distributional model that provides the closest practical approximation to the empirical data. Although several distribution families could potentially fit the observations, a meaningful characterization requires examining the lower tail, upper tail, and overall shape of the empirical distribution. Consequently, A proper approach is the Kappa family distribution.

Let $(\kappa,h)$ denote the two shape parameters, and $(\xi,\alpha)$ the location and scale parameters with $\alpha>0$. The transformed variable $t(x)$ and corresponding CDF, PDF are defined by 
(See. Appendix~\ref{appen:sec-kappa} for Kappa LMRs) 
\begin{equation}
\begin{aligned}
t(x)= &1-\frac{\kappa(x-\xi)}{\alpha} \\
F(x)= &
\begin{cases}
\exp\bigl[-t(x)^{1/\kappa}\bigr] & h\approx 0 \\
\bigl(1-h\left(t(x)^{1/\kappa}\right)\bigr)^{1/h} & \text{otherwise} 
\end{cases} \\
f(x)= &
\frac{1}{\alpha}\,t(x)^{1/\kappa- 1}\,F(x)^{\,1-h}.
\end{aligned}
\label{eq:kappa}
\end{equation}
Eq.~\eqref{eq:kappa} generalizes several critical three--parameter models. Special distributions for extreme value modeling are generalized extreme value (GEV) ($h=0$), generalized logistic (GLO) ($h=-1$), and generalized Pareto (GPA) ($h=+1$). The extra shape parameter gives Kappa an additional degree of freedom to control tail behavior independently of skewness and scale. In the LMRD, this means that the Kappa family spans a broader region of the $(\tau_3,\tau_4)$ plane and can approximate more combinations than any of the three–parameter subfamilies. Consequently, when candidate models are fitted by matching L--moments, the Kappa distribution can often move closer to the empirical point. This increased flexibility frequently leads to better empirical GoF for a particular dataset, compared to more restrictive three–parameter models.

\subsubsection{Goodness of Fit (GoF) investigation} \label{Gof}

It is expected that the Kappa distribution can achieve better GoF, but the relevant statistics must be compared to assess the rigor and avoid overfitting. Z-scores and statistical distance metrics are typically employed as defined by 
\begin{equation}
\begin{array}{c}
D^{\mathrm{dist}} = \sqrt{(\tau_{3}^{\mathrm{dist}} - \tau_{3}^{\mathrm{sample}})^{2}
      + (\tau_{4}^{\mathrm{dist}} \;-\; \tau_{4}^{\mathrm{sample}})^{2}}\\[6pt]
Z^{\mathrm{dist}} = \sigma_{\tau_4}^{-1}(\vphantom{\big|}\tau_{4}^{\mathrm{dist}} - \tau_{4}^{\mathrm{sample}}).
\end{array}
\label{eq:GOF_measures}
\end{equation}
$\tau_3^{\mathrm{sample}}$ and $\tau_4^{\mathrm{sample}}$ are extracted from the observations, and they are unbiased estimators as long as enough number of samples are available (See~\cite{silva2020moments} for other choices). In this case, the $\sigma_{\tau_4}$ can also be used as the standard deviation of the population with one DoF, while some modifications like bootstrapping or L--kurtosis bias factor compensation are required. Since Z values are basically drawn from the Gaussian distribution, a single rule of thumb for acceptance/rejection is $|Z^{\mathrm{dist}}| \le 1.54$ given a 95\% confidence interval \cite[~p.83]{hosking1997regional}. $|D^{\mathrm{dist}}|$ quantifies the geometric separation between theoretical and empirical LMRs, closer the better by definition. Joint evaluation of $|Z^{\mathrm{dist}}|$ and $|D^{\mathrm{dist}}|$, therefore, enables a reliable acceptance or rejection decision for competing distributional models. Since L--kurtosis ($\tau_{4})$ is primarily sensitive to extreme values, it reflects how well the distribution captures the heavy-tailed behavior. In contrast, D-scores jointly evaluate both $\tau_{3}$ and $\tau_{4}$ by computing the Euclidean distance between the empirical point and the closest theoretical distribution points in the LMRD, thereby quantifying overall shape consistency.


\subsubsection{Mini-summary}

L--moments and the Kappa distribution are used to obtain a robust and flexible statistical description of the empirical power distributions derived from IQ samples. Indoor multipath and obstruction effects under NLOS conditions produce heavy–tailed, skewed, and outlier–contaminated data, for which conventional (central tendency) moments and Gaussian/Rayleigh–type models can be unstable or misleading. L--moments provide scale–free measures of variability, skewness, and kurtosis that remain well defined under heavy tails, making them suitable for comparing LOS and NLOS regimes in the LMRD. The four–parameter Kappa distribution is additionally employed as a flexible parent model whose LMRs cover a wide region of LMRD and can closely match the empirical $(\tau_3,\tau_4)$ of different propagation conditions, enabling more accurate GoF assessment, parametric modeling, and downstream ML for LOS/NLOS discrimination. 

\subsection{Density-Based Spatial Clustering of Applications with Noise (DBSCAN)}
\label{sec:dbscan}

 DBSCAN is an unsupervised, density-based clustering method that partitions a dataset $\mathcal{X}=\{x_i\}_{i=1}^N\subset\mathbb{R}^d$ into clusters and a noise set by exploiting local neighborhood cardinality \cite{khan2014dbscan}. For a predefined radius $\varepsilon>0$, $\varepsilon$-neighborhood of a point $x$ is defined by 
\begin{equation}
N_\varepsilon(x):=\{y\in\mathcal{X}:\ \|y-x\|\le \varepsilon\}.
\label{eq:dbscan}
\end{equation}
A point is designated as a core point when $|N_\varepsilon(x)|$ is equal to or greater than a predefined minimum number of points, indicating that $x$ lies in a sufficiently dense region of the feature space. This neighborhood-count criterion underpins the subsequent cluster expansion mechanism, wherein clusters are grown by aggregating points that are reachable through chains of overlapping $\varepsilon$-neighborhoods rooted at core points, while points that fail to satisfy density support are retained as border samples or rejected as noise. To compare ordinary third and fourth moments with LMRs, the average silhouette score \cite{kundu} is used as an internal clustering validity index, as defined by
\begin{equation}
\mathrm{Sil.}= \frac{1}{N}\sum_{i=1}^{N} S_i = \frac{1}{N}\sum_{i=1}^{n}\frac{b(i)-a(i)}{\max\{a(i),\,b(i)\}}.
\label{eq:siluette}
\end{equation}
Eq.~\eqref{eq:siluette} quantifies how well each sample fits within its assigned cluster versus the nearest alternative cluster. For each point, $\mathrm{i}$ in the LMRD, it compares the mean intra-cluster distance, $a(i)$, to the minimum mean inter-cluster distance $b(i)$ to the closest other cluster. Given the algorithm's ability to find and generate clusters randomly and automatically, DBSCAN is employed in this paper to demonstrate the usefulness of L--moment analysis in ML tasks.

\section{Experimental Framework} \label{sec:exf}

In this section, the indoor laboratory propagation environment in which the data collection is performed is described. Then, the measurement campaign protocol is explained.

\subsection{Experimental Setup}
\begin{figure}
    \centering
    \includegraphics[width=1\columnwidth]{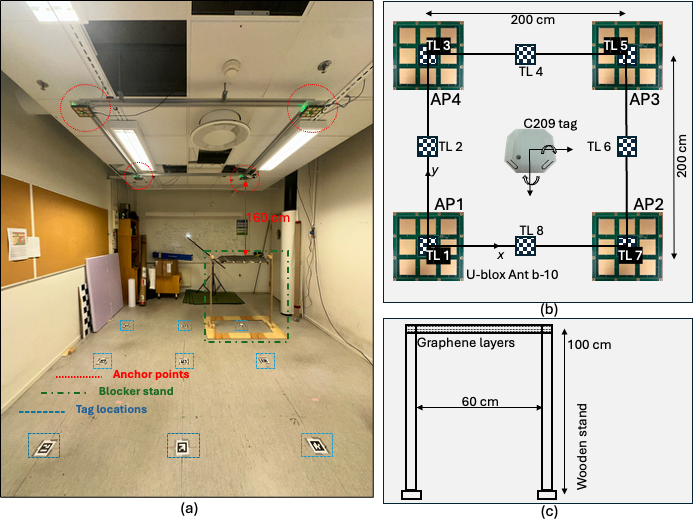}
    \caption{Experimental setup. (a) Indoor localization laboratory (b) Area of Interest (AoI) with tag location (TL) and anchor points (AP) (c) LOS blocker stand.}
    \label{fig:exprimental_setup}
\end{figure}

The experimental setup employs the u-blox XPLR-AOA-3 development kit \cite{ublox_XPLR-AOA-3_UBX-22007210_2024}, a comprehensive BLE AoA localization solution tailored for indoor positioning applications. The system is composed of ANT-B10 antenna boards functioning as anchor points, and the C209 Bluetooth tag serving as the mobile signal source. Each ANT-B10 unit integrates an eight-element patch planar antenna array and a NINA-B411 BLE 5.1 module, enabling high-resolution AoA estimation. These anchors are interfaced via Ethernet PHY and a Switch. The u-blox devices, used for experimental data collection, is presented on Fig.\ref{fig:exprimental_setup}-(b), where the IQ samples are generated and collected. The experiment was conducted in a rectangular indoor environment measuring $630\times385$\;cm, partially furnished with typical office equipment that introduces realistic multipath conditions as a prototype size of an indoor factory hall, depicted in Fig.\ref{fig:exprimental_setup}-(a). As seen, the blocker stand was positioned to obstruct the LOS propagation path between the tag—placed on the ground—and AP during an NLOS CTE measurement event; the stand was topped with a $60\times60$\;cm stack comprising layers of graphene-coated paper, thereby enforcing LOS blockage. The blocker is inserted between the anchor and tag such that it fully blocks six Fresnel zones at the carrier frequency \cite[Sec. 4.7.1]{Rappaport}, although specular reflections from walls or office objects may still reach the receiver under NLOS experiment (See. Appendix~\ref{sec-appen-F}). 

Data acquisition was performed under four measurement scenarios using a single anchor point with eight dual-polarized antenna elements mounted at a height of 260\,cm and aligned parallel to the floor, which is the typical ceiling-mounted height configuration. Four APs are mounted at the corner of the AoI, which is a 2D square of $176\times176$\,cm, but only one anchor is active in each measurement event, while the others are inactive. This ensures that all received CTE packets correspond to a single active anchor at a time, while keeping the physical array layout fixed across LOS and NLOS scenarios. The tag was deployed at eight predefined locations, within the room, as illustrated in Fig.~\ref{fig:exprimental_setup}-(b) by $\mathrm{TL} \in \{1, \cdots,8\}$. To make the case with only the NLOS signal records, the radio blocker was placed between the transmitter and the receiving antennas, one at a time. These dimension is chosen properly to emulate a realistic indoor deployment while allowing the blocker to create a distinct LOS and NLOS experiment. The blocker dimensions are depicted in Fig.~\ref{fig:exprimental_setup}-(c). The transmitter antenna (tag) was tested at three spatial directions and in three orientations. The original direction is depicted in Fig~\ref{fig:exprimental_setup}-(b), while a new direction is a 90° rotation along the xy axis. For each direction, three orientations (0°, 45°, and 90°) are tested, resulting in nine measurement configurations. Although this experimental setup increases the required resources for data collection and analysis, it increases the dataset diversity for further ML-based analysis.

\subsection{Measurement Campaign}
The measurement campaign was conducted using a single transmitter–receiver pair with both devices remaining in fixed positions during a measurement event. Data are recorded and automatically stopped for $\mathrm{N}=1000$ CTE packets. LOS and NLOS data were collected sequentially at each TL position using an identical HW configuration; for the NLOS CTE measurements, a blocker was the only variable introduced. The LOS/NLOS pairing and file labeling were assigned based on the controlled physical setup and subsequently corroborated via statistical validation. Measurement order and experimental parameters were logged and enforced by the data-collector scripts such that, for each event, the complete LOS dataset was acquired first, followed immediately by the corresponding NLOS dataset. The inter-run time gap was typically only a few minutes, and campaigns were conducted during a single working-week period to minimize uncontrolled variations in the indoor propagation environment. This protocol helped maintain a quasi-static scene across repeated acquisitions; if abrupt disturbances (e.g., human motion) occurred, the measurement was repeated to preserve environmental consistency. A set of controlled parameters is systematically varied at each measurement event to ensure diversity in the collected dataset. The transmitter antenna placement is adjusted over three discrete positions, denoted as $D \in \{1,2,3\}$. Additionally, the transmitter antenna orientation is rotated in increments of $45^{\circ}$, resulting in $\theta \in \{0^{\circ}, 45^{\circ}, 90^{\circ}\}$. The tag location is alternated between known points in the AoI, represented as $P \in \{x,y\}$. To capture both LOS and NLOS conditions, a boolean operator $M \in \{0,1\}$ is used to indicate the presence or absence of a physical blocker. The measurements are further extended across four different access points (APs), identified by $\mathrm{AP} \in \{1,2,3,4\}$ to ensure sweeping the AoI once in a single measurement campaign. For each configuration, the IQ data are collected at the CTE level based on Eq.~\eqref{eq:IQ_sample}, where $\mathbf{X}^{(n)}=(\mathbf{X}_1^{(n)}, \cdots, \mathbf{X}_{64}^{(n)})$ given the switching and timing standards for AoA \cite[p. 25]{BleOverview}, and our BLE hardware selection. Finally, the data files are randomly partitioned into training, validation, and testing datasets to avoid any accidental data leakage in further ML-based modeling and evaluation. Since the number of IQ samples for the LOS and NLOS labels is exactly the same, the training dataset is overall unbiased toward each label. However, randomization causes a small difference in the percentage of specific IQ samples chosen for the training dataset, as presented in Table~\ref{tab:dataset}. Thus, the statistical analysis in this manuscript is entirely based on the randomized training dataset.
\begin{table}[!ht]
\centering
\caption{Dataset general statistics}
\begin{tabular}{|p{65pt}|p{25pt}|p{115pt}|} \hline 
\small
 item &  value &  description  \\
 \hline
  Receiver positions & 12 & P \\
Samples per event & 1000  & N \\  
  packets per AP & 54\;K & $P\times M\times D\times\theta\times N$ \\ 
 data packets & 216\;K & \small $AP\times P\times M\times D\times\theta\times N$ \\ 
  IQ imaginary values & 15,552\,K & \small $IQ \times AP\times P\times M\times D\times\theta\times N$ \\ 
   Data files & 216 & \small $AP\times M\times \theta \times D \times P$ \\  
    Training samples & 132\;K & 66+66 (data files) \\ 
   Validation samples & 42\;K & 21+21 (data files) \\
    Testing samples & 42\;K & 21+21 (data files)  \\
        LOS samples & 65\;K & $\approx 49\%$  \\
               NLOS samples & 67\;K & $\approx 51\%$  \\
               CTE length & 64 & Number of IQ samples $(\mathbf{X^{(n)}})$\\
               Slot duration & 1 & $\mu$\;sec.\\
               Switching duration & 1 & $\mu$\;sec.\\
               TX power & 0 & dBm\\\hline
\end{tabular}
\label{tab:dataset}
\end{table}

\section{Experimental results}
\label{sec:ExpResults}

This section presents a stepwise analysis pipeline designed to verify that the collected BLE CTE measurements for LOS and NLOS conditions have a clear statistical signature at the IQ absolute power level. First, mean and representative exemplars for each class are extracted by computing column-wise means of the training CTE packets. Second, to support rigorous hypothesis testing on class-wise averages, outliers are removed using predefined filtering rules, given the fact that average value analysis of the dataset is outlier sensitive. Hypothesis tests, then, are applied feature-wise to ensure that first-moment differences are statistically significant. Third, an L--moment-based analysis is carried out to characterize higher-order distributional properties of the IQ power features labeled by LOS/NLOS during the measurement campaign. A bootstrapped mean clouds for each label are then fitted to the empirical Kappa distribution curves, showing that Kappa-based distributional modeling is more effective than theoretical counterparts. Finally, Goodness of Fit (GoF) metrics in the LMRD are presented and analyzed. 

\subsection{Ordinary Mean and Exemplar Comparison}\label{examplar}
\begin{figure}[!ht]
    \centering
    \includegraphics[width=1\columnwidth]{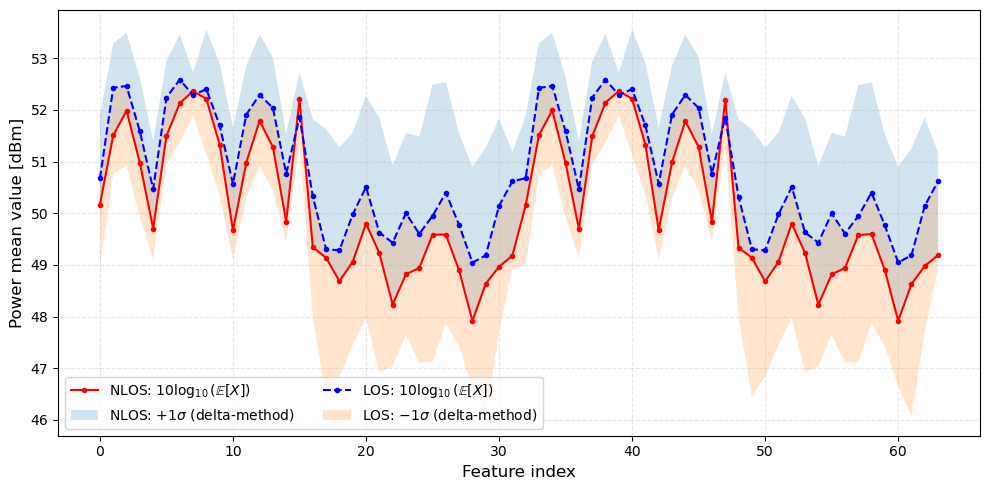}
    \caption{Raw feature power moments}
    \label{fig:ordinary-means}
\end{figure}

\begin{figure}[!ht]
    \centering
    \includegraphics[width=1\columnwidth]{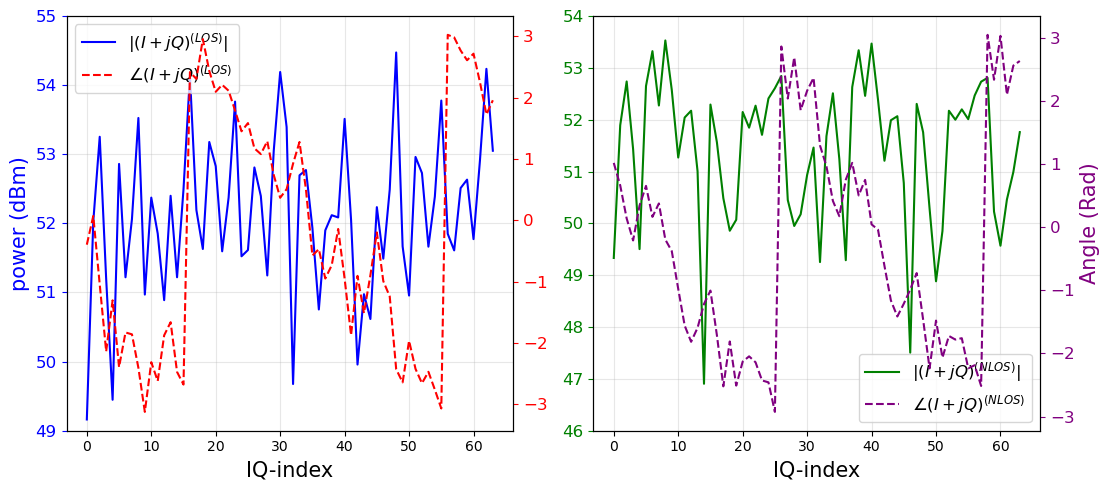}
    \caption{Sample exemplars based on mean values.}
    \label{fig:LOS_NLOS_exemplars}
\end{figure}

As an initial analysis, the column-wise ordinary power means of the training CTE packets are computed, where mW to dB scale transform is performed after mean calculation and variances are computed based on delta method, which is the Taylor expansion to approximate a random vector \cite[p. 25]{van2000asymptotic}. As observed in Fig.~\ref{fig:ordinary-means}, the ordinary (product-moment) means are often closely spaced between LOS and NLOS, and in some cases the NLOS power mean exceeds the LOS mean. This behavior is physically consistent with multipath propagation, where the superposition of signal replicas with favorable relative phases can yield constructive interference, thereby increasing the average received power despite the absence of a direct path.
For each class, the medoids are subsequently determined by computing the Euclidean distance between each received signal vector $\mathbf{u} \in \mathbb{R}^{1 \times 64}$ and the corresponding column-wise mean vector. The index yielding the minimum total distance is selected as the most representative exemplar for that class. The result for a single row index that has the closest distance to the mean values is presented in Fig.~\ref{fig:LOS_NLOS_exemplars} as two LOS and NLOS exemplars.
As observed in Fig.~\ref{fig:LOS_NLOS_exemplars}, the maximum received signal power for the LOS and NLOS class are close to each other, however, the LOS range of variation is less than NLOS exemplar. The reduced power observed in the NLOS case and higher abruption is consistent with the attenuating and diffusive propagation effects introduced by the graphene layers acting as an obstructor and reflector, particularly since all other experimental parameters were held constant throughout the measurements. In contrast, the phase trace of the NLOS exemplar appears more irregular and unstructured variation, indicating increased multipath dispersion and reduced phase coherence under NLOS conditions. These observations collectively signal the discriminative LOS/NLOS subset characteristics.

\subsection{Outlier Detection and Removal}
\label{sec:outlier_results}

Since the employed statistical tests rely on class-wise average and variance values, as mentioned in \ref{subsec:statistical_tests}, outlier detection and removal are essential. To address this, three filtering rules are introduced in Section~\ref{subsec:outlier_detector}. Design parameters for MAD are chosen based on \cite{yaro2023outlier},   $\alpha=1.4826$ and $\tau=1.28$, which is derived based on several BLE RSSI simulation campaigns and should be viewed as tuned parameters for this environment rather than an exact Gaussian consistency factor. Given the number of observations and features, the support fraction and chi-squared value for approximation is configured to $\mathrm{h-subset}=0.62$ and $\chi^2 =0.95$, respectively for outlier detection. The results are depicted in Table~\ref{tab:outlier_reduction}. Combined application of outlier detectors to the training dataset yields a refined set of 12{,}007 samples out of the original 132{,}000 ones. As discussed previously, each outlier detector targets a different criterion as an indicator of anomalous behavior.
\begin{table}[!ht]
\centering
\small
\caption{CTEs reduction rate by using different outlier detectors}
\begin{tabular}{|c|c|c|c|c|} \hline 
 IQR &  MAD &  MCD & Union & Intersection  \\\hline
 32.6 & 89.9 & 56.6 & 90.9 & 30.6 \\\hline
\end{tabular}
\label{tab:outlier_reduction}
\end{table}
For the subsequent analysis, all decision rules are implemented conservatively to retain the observations closest to the central tendency within the resulting subsets, even though this reduces the overall sample size. Afterward, the LOS and NLOS labels is assigned to the remained CTE packets. If, under these harsh conditions, a statistically significant difference between the LOS and NLOS statistical characteristics is still observed, this provides strong empirical evidence that the experiment remains sufficiently rich to generate LOS/NLOS distinction.

\subsection{Statistical Hypothesis Testing}
\label{sec:stat_results}
The statistical analyses described in Section~\ref{subsec:statistical_tests} are then applied to cleaned dataset, and the resulting mean and variance differences between LOS and NLOS samples are extracted. 

\begin{figure}[!ht]
\centering
\subfloat[Welch $t$-statistics.]{%
  \includegraphics[width=0.49\columnwidth]{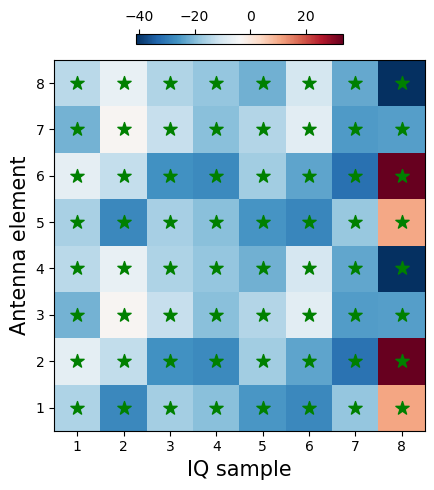}%
  \label{fig:stat_test_welch}%
}\hfill
\subfloat[Levene $F$-statistics.]{%
  \includegraphics[width=0.49\columnwidth]{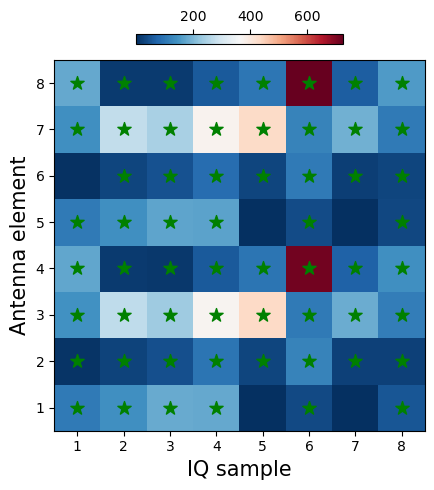}%
  \label{fig:stat_test_levene}%
}
\caption{Statistical tests of moment equality ($\alpha \le 0.01$).}
\label{fig:stat_test}
\end{figure}
Given the antenna layout of the BLE AoA receiver, the 64D feature space is reshaped into an $8\times 8$ pixel image to facilitate spatial interpretation of the statistical hypothesis tests on mean and variance differences. As illustrated in Fig.~\ref{fig:stat_test}, all $t$-statistics exceed the theoretical counterparts within $\alpha\le 0.01$ threshold, indicating that, with 99\% confidence, the class-wise mean differences across all antenna–sample index pairs are statistically significant, and the null hypothesis of equal means can be rejected. The adjusted $p$-values control for all pair comparisons, according to BH-FDR procedure, are smaller than $1e^{-6}$, demonstrating the excellent LOS/NLOS separability. Moreover, BH-FDR confirms that minor modifications do not change the statistical significance. This is visually highlighted through asterisk markers denoting significant entries. For the best IQ-antenna pairs in terms of t-statistics, the mean difference is approximately 40\,mW in received power between LOS and NLOS scenarios, while the worst IQ-antenna pairs have still around 5\;mW difference. Furthermore, the alternating sign of t-statistics confirms constructive/destructive multipath effects across the antenna–sample. A similar analysis of the variance differences shows that most feature pairs also exhibit statistically significant differences, except for five antenna-IQ-sample pairs. The worst-case scenario belongs to the first IQ sample obtained from the sixth antenna element with a $p$-value of 0.65. The best case scenario among these insignificant values belongs to the fifth IQ sample acquired from the first antenna element with the $p$-value of 0.015. The mean NLOS received power is higher than the LOS counterpart just for one out of five insignificant IQ-antenna pairs. As seen, there is still other IQ samples on the same antenna element with significant $p$-values and vice versa. Similarly, BH-FDR control verifies highly separable LOS/NLOS regime (the worst significant adjusted $p$-value is in the order of $10^{-3}$). Thus, given that the class-wise expected differences are statistically significant across nearly all dimensions, the results apparently indicate that the two classes correspond to distinct underlying physical propagation mechanisms and distributional characteristics.  

\subsection{L--moments Analysis}
\label{l-moment_analysis}

It is crucial to verify that the dataset is appropriate for L--moment analysis. This involves ensuring that all observations faithfully represent the measured quantity and originate from a common underlying distribution, which should be confirmed through an initial screening step. It includes identifying locations that are grossly discordant with the group mean, where discordancy is assessed using the LMRs. This analysis can be considered both as a quick inspection to diagnose or an initial step for outlier removal scale-free variability, asymmetry, tail-heaviness, and potential outliers/heterogeneity before committing to a parametric model \cite{hosking1997regional}. Next, we use the LMRD by plotting the empirical point ($\tau_3,\tau_4$) against theoretical distribution of candidate families; this step is an efficient model screening and selection because many distributions occupy distinct regions in the LMRD, so incompatible models can be rejected visually and analytically. After narrowing candidates, we fit a flexible family such as the Kappa distribution by solving nonlinear minimization equations that equate theoretical ratios to their sample estimates, yielding parameter estimates. Finally, we validate the fitted model via GoF diagnostics to quantify whether discrepancies are consistent with sampling variability. Skipping each step risks fitting an overly flexible model to incompatible LMRs, masking heterogeneity/outliers, producing poorly identified parameters, and ultimately passing a model that explains the data numerically but not structurally.

\subsubsection{Discordancy Analysis}\label{discordancy}

The first two LMRs (L--CV, L--skewness) of a single feature column can be viewed as points in a three-dimensional feature space, forming a cloud over all features, both for LOS and NLOS experiments. A column is flagged as discordant if its corresponding point lies far from the center of this cloud, where distance is measured by a Mahalanobis metric that accounts for the correlation among the L--moment ratios (cf.~\eqref{eq:mahalanobis}). In practice, the Mahalanobis distance is computed in the L--skewness-L--CV plane. First, a mean L-skewness--CV vector is estimated over all feature columns, and second, the discordancy ellipses in the diagram are obtained as contours of constant Mahalanobis distance between each feature column and this central point. Therefore, the heuristic aim is to identify features (columns) whose L--moment signatures deviate significantly from the overall pattern.
\begin{figure}[!ht]
\centering
\subfloat[NLOS]{%
  \includegraphics[width=0.49\columnwidth]{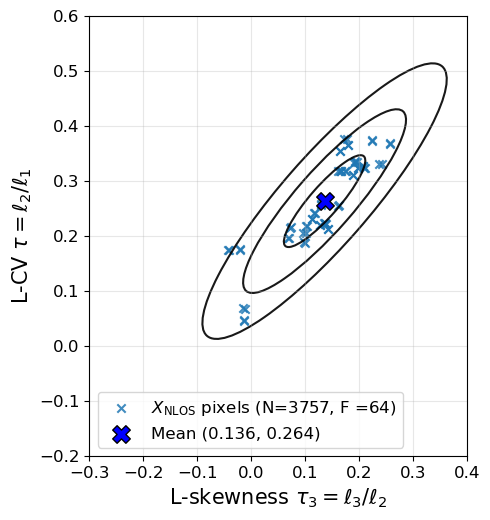}%
  \label{fig:cv_los}%
}\hfill
\subfloat[LOS]{%
  \includegraphics[width=0.49\columnwidth]{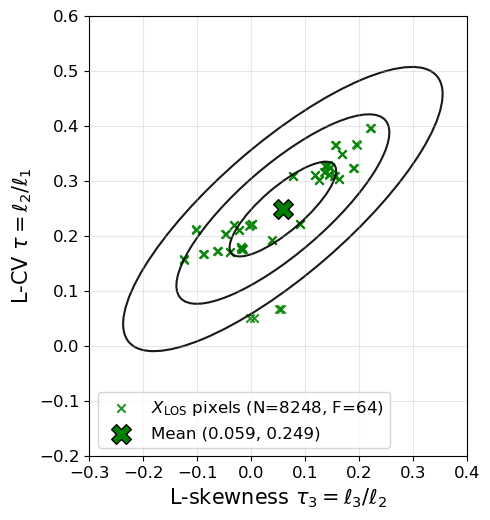}%
  \label{fig:cv_nlos}%
}
\caption{Discordancy diagrams for LOS and NLOS feature sets.}
\label{fig:l-mmoments-CV}
\end{figure}
As shown in Fig.~\ref{fig:l-mmoments-CV}, the discordant feature columns appear outside the second concentric ellipse, which represents the correlation-adjusted confidence boundary for the LMRs. The NLOS L--skewness values exhibit a right-skewed pattern, indicating a tendency toward heavier upper tails in the corresponding empirical distributions, compared to LOS L--skewness. Moreover, the ratio of the minor to major principal axes of the fitted ellipses is larger for the LOS class, which implies that the LOS L--moment pairs are more isotropic and tightly clustered. Thus, the variability is more evenly distributed in all directions of that 2D L--moment spaces, and there is less pronounced stretching along a single dominant principal direction. In contrast, the more elongated NLOS ellipses indicate stronger anisotropy and correlation between the L--moment ratios, with substantially larger spread along one principal axis. Physically, this is consistent with LOS propagation being more homogeneous and stable across feature space, whereas NLOS propagation exhibits more heterogeneous, directionally structured variability due to complex obstruction effects. Based on the discordancy analysis, column features whose L--moment coordinates fall outside the second concentric ellipse are treated as feature-wise outlier dimensions and subsequently removed from the dataset. Applying this criterion to both LOS and NLOS measurements results in a reduced and more homogeneous feature set, leaving 40 and 42 out of 64 column features for LOS and NLOS, respectively. Removed features belong to both the IQ and antenna elements axis on several switching instances, indicating that several sources of deviation are excluded from each subset. Thus, applying this removal satisfies the concordant requirements for the subsequent analysis. From a physical standpoint, the adopted 8-element planar antenna array exhibits mutual coupling and induces spatial correlation across the antenna–sample IQ feature grid. Therefore, applying feature-wise (marginal) outlier filtering constitutes a conservative preprocessing step: it suppresses per-feature extremes without leveraging the underlying cross-feature dependence, while retaining the full $8\time8$ structure for subsequent multivariate analysis.
\subsubsection{L--Skewness--L--kurtosis Bootstrapping}\label{bootstraping}
\begin{figure}[!ht]
    \includegraphics[width=1\columnwidth]{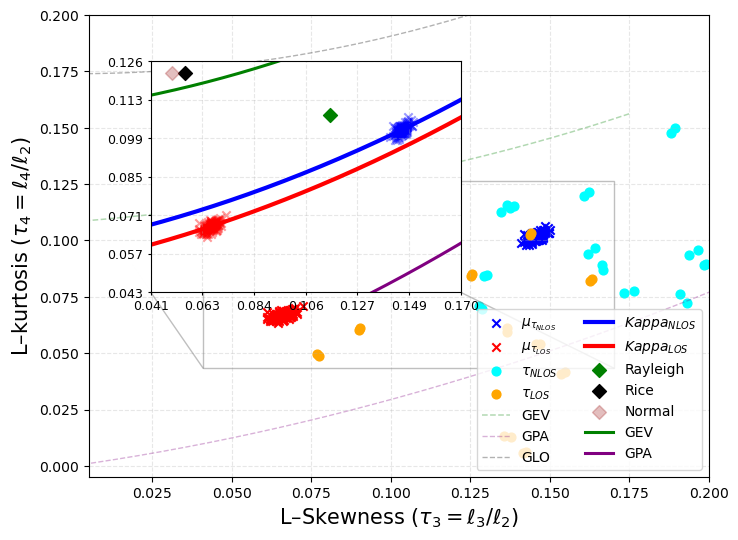}
    \caption{LOS and NLOS L--moments ratio diagram (LMRD).}
    \label{fig:tau3tau4}
\end{figure}

To ensure that the L--skewness--L--kurtosis characteristics of the LOS/NLOS datasets are statistically comparable, a nonparametric bootstrapping scheme is employed. Blocks of 120 subsamples are repeatedly drawn, and the corresponding mean values of $\tau_{3}$ and $\tau_{4}$ are computed. These bootstrap averages, represented in red and blue clouds in Fig.~\ref{fig:tau3tau4}, exhibit substantially reduced variability compared with the original column-wise estimates, yielding clusters of mean-of-means that lie in close proximity, as expected for internally homogeneous samples. 
A magnified view of this region is then examined for further discrimination. The NLOS kurtosis cloud clearly lies above the LOS cloud, indicating heavier tails driven by more frequent extreme fades of the channel. Moreover, both LOS and NLOS exhibit positive skewness, implying a systematic asymmetry that suggests theoretical models such as the Rician distribution are not realistic approximations in practice \footnote{To see how LMRs are extracted for Rayleigh, Rice and Two-parameter Normal distributions, ref. Appedix~\ref{sec:appen-distro}. Theoretical two-parameter distributions correspond to a single point in the LMRD because their LMRs depend only on a single distributional shape. Moreover, the associated L--moment analysis for all distributions is intrinsically scale invariant.}

\subsubsection{Kappa Distribution Identification}\label{kappa fitting}
To determine the optimal Kappa shape parameters $(\kappa,h)$ for both LOS and NLOS datasets, a numerical optimization procedure is carried out using a least–squares objective defined on the LMRD (See.~\cite{khan2025comparative}). The objective function evaluates the discrepancy between the model-implied L--skewness-L--kurtosis of the Kappa distribution and their target bootstrap means as defined by 
\begin{equation}
J(\kappa,h)=\bigl(\tau_{3}^{\text{model}}(k,h)-\tau_{3}^{\text{ref}}\bigr)^{2}
       +\bigl(\tau_{4}^{\text{model}}(k,h)-\tau_{4}^{\text{ref}}\bigr)^{2}.
\label{eq:kstar}
\end{equation}
with implicit constraints $\lvert k\rvert<0.99$ and $\lvert h\rvert<0.99$ to ensure admissible Kappa forms, the minimization of $J(k,h)$ is performed separately for LOS and NLOS by initializing at $(0.1,0.1)$ and applying a bounded quasi-Newton search, yielding the optimal parameters $(\kappa^{\ast},h^{\ast})^{\mathrm{NLOS}} = (0.2, 0.44)$ and $(\kappa^{\ast},h^{\ast})^{\mathrm{LOS}} = (0.42, 0.51)$. After obtaining the optimal parameters, the corresponding L--moment equations are used to reconstruct the Kappa curves for both propagation conditions. As illustrated in Fig.~\ref{fig:tau3tau4}, the resulting curves intersect the bootstrapped $(\tau_{3},\tau_{4})$ clusters with high precision, validating the fitted shape factors. 
Although the averaged $(\tau_{3},\tau_{4})$ values for LOS and NLOS appear close, they form distinguishable groupings when contrasted with theoretical distributional curves in the LMRD and when incorporating the fitted shape parameters of the Kappa family distribution. This separation provides evidence that the underlying generative mechanisms for LOS and NLOS differ in higher-order L--moment structure despite similar ordinary central tendencies as depicted in Fig.~\ref{fig:ordinary-means}. Moreover, the NLOS bootstrap cloud leaning toward from Kappa to the theoretical generalized extreme-value (GEV) trajectory, which is a special case of Kappa family ($\mathrm{h}=0$), whereas the LOS cluster is positioned closer to the generalized Pareto (GPA) curve, as expected, which is a special case of Kappa distribution ($\mathrm{h}=1$). From the physical perspective, the elevated L--kurtosis and heavier tails observed in the NLOS subset are consistent with a rich multipath propagation regime, where constructive and deconstructive interference generates both deep fades and occasional strong received power peaks. These effects naturally embed in skewed, heavy-tailed distributions that are poorly represented by light-tailed Gaussian or Rayleigh theoretical models. In contrast, the LOS subset exhibits lower L--kurtosis, while still remaining mildly non-Gaussian, reflecting a dominant direct path superimposed with weaker reflective or scattering paths.

\subsubsection{GoF Statistics}\label{GoF_stats}
To investigate GoF, using the closest point on each theoretical LMRD curve is preferable because it enforces geometric consistency, i.e., the selected ($\tau_3,\tau_4$) pair corresponds to a single admissible parameter value on the curve, rather than combining L--skewness-L--kurtosis from different indices, which can yield an unattainable “hybrid” point and artificially deflate the GoF distance. Under this proper 2D projection, the GPA family typically loses its apparent advantage because its curve is comparatively steep and strongly parameter-coupled in the neighborhood of the empirical centroid; hence the nearest feasible point on the distribution locus is farther in Euclidean ($\tau_3,\tau_4$) distance, especially reducing the observed GPA GoF. Z and D scores are computed after this modification for the candidate distributions based on Section~\ref{Gof} and summarized in Table~\ref{tab:gof_results}. 
\begin{table}[h!]
\centering
\caption{GOF metrics: Z-statistic and Euclidean distance.}
\label{tab:gof_results}
\begin{tabular}{|l|l|l|l|l|} \hline
 & $Z_{\mathrm{NLOS}}$ & $D_{\mathrm{NLOS}}$ & $Z_{\mathrm{LOS}}$ & $D_{\mathrm{LOS}}$ \\ \hline
$\mathrm{Kappa_{NLOS}}$ & 0.011809 & 0.000320 & -- & --\\
$\mathrm{Kappa_{LOS}}$ & -- & -- & 0.000177 & 0.000171\\
Rayleigh & 0.245754 & 0.030598 & 1.409236 & 0.063921 \\
GEV & 1.745148 & 0.038999 & 1.753490 & 0.051161 \\
GPA & 2.020198 & 0.047646 & 1.556611 & 0.047049 \\
GLO & 4.424334 & 0.099222 & 3.892981 & 0.112401 \\
Rice & 1.094312 & 0.147427 & 2.033890 & 0.087712 \\
Norm & 1.096798 & 0.147787 & 2.035721 & 0.088014 \\\hline
\end{tabular}
\end{table}
As expected, the Kappa family provides the closest L--moment match to the empirical channel statistics when each class is modeled with its own fitted parameters. Specifically, $\mathrm{Kappa_{NLOS}}$ attains the smallest discrepancy in the LMRD with near-zero standardized L-kurtosis deviation, while $\mathrm{Kappa_{LOS}}$) yields an even tighter LOS fit. In contrast, classical flat-fading baselines and common extreme-value families exhibit orders-of-magnitude larger Euclidean offsets and substantially larger D, indicating systematic mismatch in tail behavior and/or asymmetry. Overall, the results support that propagation-conditioned Kappa modeling captures the heavy-tailed NLOS statistics. Additionally, It matches to the more concentrated LOS regime substantially better than conventional distributions for flat fading channels.


\subsection{DBSCAN Clustering}

\begin{figure}[!ht]
    \centering
    \includegraphics[width=1\columnwidth]{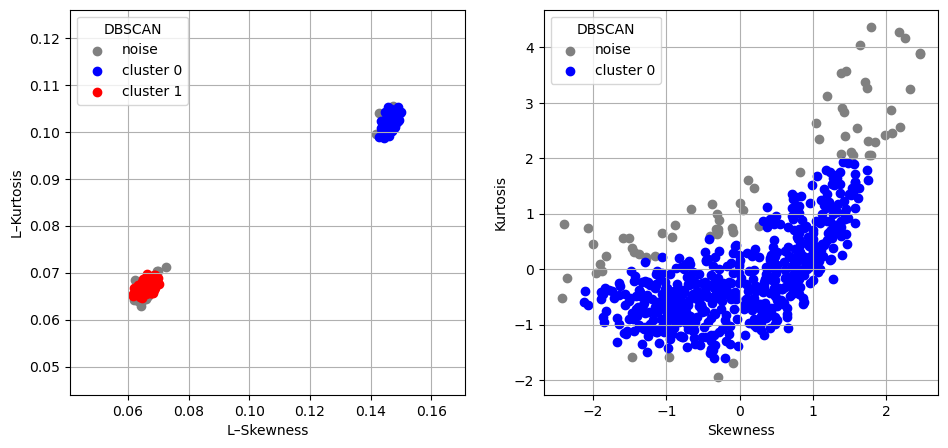}
    \caption{DBSCAN clusters}
    \label{fig:DBSCAN}
\end{figure}

To demonstrate the utility of L--moments as feature descriptors for ML tasks, we generate bootstrap replicates by partitioning the measurements into $10$ blocks and computing both ordinary (product) moments and L--moment-based statistics for each block. Given 64 IQ samples, a sample-feature matrices of $640\times4$ is generated, where the rows are bootstrapped IQ features and the columns are ordinary integer moments. The resulting feature matrices are then standardized and provided as inputs to the DBSCAN clustering algorithm. The clustering outcomes are depicted in Fig.~\ref{fig:DBSCAN}. As expected, the LMRs yield two well-separated clusters that are consistent with the underlying LOS and NLOS regimes, whereas the ordinary-moments collapse predominantly into a single cluster.
Using DBSCAN internal validation based on the Silhouette score (computed on non-noise points), the L--moment feature space $(\tau_3,\tau_4)$ yields two nontrivial clusters with zero noise ($\mathrm{N_{LOS}}=368, \mathrm{N}_{NLOS}=272$), null noise fraction, and a measurable, albeit small, separation quantified by ($\mathrm{Sil.}=0.018)$). In contrast, clustering on ordinary product-moment features collapses to a single detected cluster ($\mathrm{N}=570$) while assigning 10.94\% of samples to noise, which makes the silhouette score undefined because silhouette requires at least two clusters. Therefore, under the same DBSCAN selection strategy, L--moment statistics reveal a multi-cluster structure in the data that is not recoverable from ordinary moments, indicating improved sensitivity of the L--moment representation to latent distributional heterogeneity. These results indicate that L--moment descriptors provide improved geometric separability in the induced feature space, which is desirable for downstream ML classification. In particular, since L--moments are linear functionals of order statistics, they exhibit reduced sensitivity to outliers and heavy-tailed behavior compared to product moments, thereby enhancing robustness under multipath-affected and non-Gaussian measurement conditions. Consequently, the observed DBSCAN separation supports the interpretation that L--moment features constitute more informative and discriminative inputs for ML tasks than ordinary moments.

\subsection{Mini-summary}
The proposed procedure provides a coherent statistical validation of LOS/NLOS separability in BLE CTE IQ measurements. Exemplar observations indicate that LOS received power range is higher with smoother phase behavior, whereas NLOS captures show stronger attenuation and irregular phase fluctuations, indicating multipath-induced degradation. Feature-wise hypothesis tests over the reshaped antenna–sample grid confirm that LOS and NLOS differ significantly in both mean and variance at for nearly all dimensions, with the majority of hypotheses yielding effectively zero adjusted $p$-values, thereby establishing distinct first- and second-order statistics. LMRD and Kappa-family fitting to bootstrapped pairs yields clearly separated LOS and NLOS clusters in the LMRD with different optimal shape parameters, demonstrating systematic differences in higher-order distributional properties. These results further indicate that Kappa-based modeling better captures tail behavior and asymmetry than standard Rayleigh/Rice assumptions. 


\section{discussion}

Based on current observation and results, we highlight both the constraints and broader implications for system design and ML–driven processing as follows.

\begin{itemize}
\item From an engineering perspective, the observed heavy-tailed NLOS power statistics indicate that classical Gaussian or Rayleigh assumptions are often inadequate for AoA LOS/NLOS channel modeling. Kappa-family models offer a more realistic stress-testing framework, while LMRs provide a meaningful geometric representation in the $(\tau_3,\tau_4)$ space. This geometry is directly linked to distributional shape differences, making it particularly well-suited for unsupervised ML methods that rely on distance metrics, such as clustering or anomaly detection.
\item The results suggest that L--moment–based features provide compact, low-dimensional descriptors that are more robust to outliers and heavy-tailed behavior than conventional product moments or raw RSSI statistics. Integrating these features into explicit LOS/NLOS classifiers and benchmarking them against baseline feature sets represents a generic next step. Such comparisons will clarify the practical gains achievable through L--moment–based preprocessing in supervised and unsupervised learning.
\item Our experiments focus on statistically stationary conditions in which flat-fading multipath effects dominate. Although this setup captures essential distributional properties, it does not fully reflect realistic indoor scenarios involving moving objects or users. Extending the framework to dynamic environments will be critical for evaluating how robust L--moment descriptors remain under time-varying multipath and intermittent shadowing effects commonly encountered in practical deployments.
\item The present analysis is limited to a single indoor environment, one BLE AoA platform, and fixed antenna array geometry and tag height. Consequently, the extracted Kappa parameters and L--moment ratio locations are inherently dependent on the specific environment and hardware configuration. While this does not diminish the validity of the statistical framework, it motivates future measurement campaigns across diverse rooms, building materials, and array layouts to assess the universality and transferability of the observed L--moment/Kappa relationships.
\end{itemize}

\section{Conclusion}
This paper presents a dedicated experimental and statistical framework for characterizing BLE AoA measurements under LOS and NLOS propagation conditions. Using a controlled indoor setup, a large-scale CTE-based dataset was collected and systematically refined via robust outlier detection. Subsequent hypothesis tests confirmed statistically significant differences in the first-order and second-order product-moments between LOS and NLOS classes, while L--moment analysis provided a robust distributional characterization in terms of L--skewness and L--kurtosis. By fitting flexible Kappa models, the underlying channel behavior was shown to be well captured, offering the best overall agreement according to GoF metrics. Since our novel LMRD/Kappa descriptors present compact, robust features extracted from a realistic LOS/NLOS channel, it can be feasibly employed to improve channel characterization and a proper option for ML tasks versus conventional product-moment representations. 




\appendices \label{sec:appen}
\section{L--moment estimation}\label{sec:appen-A}

To estimate L--moments from data $\{x_{1:n}, \cdots, x_{n:n}\}$, one begins with the unbiased estimators of probability-weighted moments (PWMs). The PWM estimator of order $r$, which coincides with the standard definition of PWMs is given by 
\begin{equation}
\widehat{\beta}_r
\;=\;
\frac{1}{n}\binom{n-1}{r}^{-1}\sum_{j=r+1}^{n}
\binom{j-1}{r}\; x_{(j:n)}.
\label{eq:beta_hat}
\end{equation}
Comparing the \eqref{eq:AlphaBeta} and \eqref{eq:beta_hat} shows that the L--moment estimator of $\ell_r$ can be extracted 
\begin{equation}
    \label{eq:PWM-f}
    \ell_r = \frac{1}{n}\sum_{j=1}^{n}w_{j:n}^{(r)}x_{j:n} .
\end{equation}
The weights, $w_{j:n}^{(r)}=(-1)^{r-1}P_{r-1}(j-1,n-1)$ resembles the polynomials of degree $r-1$ in j, analogously to the discrete Legendre polynomial as defined by probability weighted moments of \eqref{eq:lmr-definition} and \eqref{eq:AlphaBeta}.
\section{L--moment ratio derivatives}\label{sec:appen:ratios}
Using \eqref{eq:LegendrePolinomials}, gives the L--moment coefficients directly: 
\begin{equation}
\begin{aligned}
P_0^*(u) &= 1, \\[6pt]
P_1^*(u) &= 2u - 1, \\[6pt]
P_2^*(u) &= 6u^{2} - 6u + 1, \\[6pt]
P_3^*(u) &= 20u^{3} - 30u^{2} + 12u - 1.
\end{aligned}
\label{eq:polyCoeff}
\end{equation}
Using \eqref{eq:lmr-definition} and \eqref{eq:AlphaBeta} yields the first four L--moments as 
\begin{equation}
\begin{aligned}
\lambda_1 
&= \int_{0}^{1} Q(u)\, 1\, du 
= \beta_0 \\[4pt]
\lambda_2 
&= \int_{0}^{1} Q(u)\,(2u - 1)\,du 
= 2\beta_1 - \beta_0 \\[4pt]
\lambda_3 
&= \int_{0}^{1} Q(u)\,(6u^{2} - 6u + 1)\,du \\[4pt]
&= 6\beta_2 - 6\beta_1 + \beta_0 \\[4pt]
\lambda_4 
&= \int_{0}^{1} Q(u)\,(20u^{3} - 30u^{2} + 12u - 1)\,du \\[4pt]
&= 20\beta_3 - 30\beta_2 + 12\beta_1 - \beta_0.
\end{aligned}
\label{eq:lambdafour}
\end{equation}
Finally, the L--CV, L--skewness, and L--kurtosis is given by 
\begin{equation}
\tau_1 = \frac{\lambda_2}{\lambda_1},\quad
\tau_r = \frac{\lambda_{2r+1}}{\lambda_{2r}},\ r\ge 1,\quad
\rho_r = \frac{\lambda_{2r}}{\lambda_{r}},\ r\ge 2.
\label{eq:taufour}
\end{equation}
(Ref.~\cite{alvarez2022inference} to see generalized higher L--moment estimators). 

\section{L--moments of particular distributions}\label{sec:appen-distro}
\textit{L--moments of two-parameter Normal distribution:} Let $Z\sim\mathcal{N}(0,1)$ and $X = \mu +\sigma Z$. By linearity of L--moments under location-scale transforms, $\lambda_1(X) = \mu +\sigma\lambda_1(Z)$ and $\lambda_r(X) = \sigma\lambda_r(Z)$. Thus for a standard Normal $Z\sim\mathcal{N}(0,1)$, with the CDF $\Phi(z)$ and PDF of $\varphi(z)$, set $u=\Phi(z)$, so $du=\varphi(z)dz$. Then, define the L--moments  
\begin{equation}
\begin{aligned}
\lambda_r &= \int_{0}^{1} Q(u)\,P_{r-1}^*(u)\,du = \int_{-\infty}^{\infty} z\,P_{r-1}^*\left(\Phi(z)\right)\varphi(z)dz.\\
\lambda_1 &= 0\quad
\lambda_2 = \frac{1}{\sqrt{\pi}}\quad
\lambda_3 = 0\quad
\lambda_4 \approx 0.0691706.
\end{aligned}
\label{eq:normMomentVal}
\end{equation}
\textit{L--moments of Rayleigh distribution}:
Let $X\sim\mathrm{Rayleigh}(\sigma)$. The CDF and quantile function is obtained by solving $u=F(x)$, as defined by 
\begin{equation}
\begin{aligned}
F(x) &= 1 - \exp\!\left(-\frac{x^{2}}{2\sigma^{2}}\right) 
\qquad x\ge 0 \\[4pt]
Q(u) &= \sigma\sqrt{-2\ln(1-u)} 
\qquad 0<u<1.
\end{aligned}
\label{eq:RayleighQ}
\end{equation}
Given \eqref{eq:AlphaBeta}, introduce the change of variable $t=-\ln(1-u)$, i.e.\ $u=1-e^{-t}$ and $du=e^{-t}\,dt$, using the binomial expansion of $(1-e^{-t})^r$ and Gamma integral, yields  the closed form solution for $\beta$ function. Pluging in \eqref{eq:lambdafour}, gives the final values: 
\begin{equation}
\begin{aligned}
\beta_r &= \sigma\sqrt{\frac{\pi}{2}}\sum_{k=0}^{r}(-1)^{k}\binom{r}{k}(k+1)^{-3/2}\\
\lambda_1 &\approx 1.25\,\sigma\quad
\lambda_2 \approx 0.38\,\sigma\quad
\lambda_3 \approx 0.042\,\sigma\quad
\lambda_4 \approx 0.039\,\sigma.
\end{aligned}
\label{eq:RayleighBetaGeneral}
\end{equation}
\textit{L--moments of rice distribution}:
Let $X \sim \mathrm{Rice}(\nu,\sigma)$ with the PDF and CDF of $f(x)$ and $F(x)$, respectively. Given the change of variable $u = F(x), \quad Q(u)=x, \quad du = f(x)\,dx$, The (r)-th L--moment is defined in quantile form as \eqref{eq:lmr-definition}. Substituting gives \eqref{eq:rice_general}, which does not have a closed-form solution, but it can be solved numerically. For  $v=8$ and $\sigma=0.5$, corresponding to Rician K-factor $K=v^2 / 2\sigma^2=128$ or $K=21.1$\;dB as a strong LOS power, the numerical L--moments are the following:
\begin{equation}
\begin{aligned}
\lambda_r
&= \int_{0}^{\infty} x\,P_{r-1}^*\!\bigl(F(x)\bigr)\,f(x)\,dx
\qquad r\ge1\\
\lambda_1 &=\approx 8.016\quad 
\lambda_2 \approx 0.281 \quad
\lambda_3 \approx 0 \quad
\lambda_4 \approx 0.0345.
\end{aligned}
\label{eq:rice_general}
\end{equation}

\section{Kappa distribution and its L--moment}\label{appen:sec-kappa}
Kappa distribution is presented in \eqref{kappa}. For $u\in(0,1)$, the quantile function, $Q(u)$, is given by 
\begin{equation}
\begin{aligned}
A(u;h) &=
\begin{cases}
-\ln u, & h\to 0 \\
\dfrac{1-u^{h}}{h}, & h\neq 0.
\end{cases} \\
Q(u) &=
\begin{cases}
\xi-\alpha\ln A(u;h), & k\to 0 \\
\xi + \dfrac{\alpha}{\kappa}\bigl(1-A(u;h)^{\,\kappa}\bigr) & \text{otherwise.}
\end{cases}
\label{eq:kappaQuantile}
\end{aligned}
\end{equation}
Configuring the shape parameters $(\kappa,h)$ in the Kappa family enables the construction of several critical three-parameter distributions. 
The availability of a closed-form solution for $Q(u)$ enables numerical L--moments computation using \eqref{eq:lmr-definition} and \eqref{eq:LegendrePolinomials} or a theoretical solution for predefined distributions (i.e., known shape parameter of $\kappa$ and h) using \eqref{eq:AlphaBeta}. The second option gives $\beta_r \in \mathbb{Z} \ge 0 $ as a function of Kappa shape parameters as follows: 
\begin{equation} 
\beta_r =
\frac{\xi}{r+1} +
\frac{\alpha}{\kappa}
\left[
\frac{1}{r+1}-\frac{1}{h^{\,\kappa+1}}\,\beta \!\left(\frac{r+1}{h},\,\kappa+1\right)\right].
\end{equation}
Having the beta function gives the moments as a function of shape factors, as defined by 
\begin{equation}
\begin{aligned}
\lambda_1 & = \beta_0 = \xi + \frac{\alpha}{\kappa}\Bigl(1 - \beta_1\Bigr) \\
\lambda_2 & = 2\beta_1-\beta_0 = \frac{\alpha}{\kappa}\Bigl(\beta_1 - \beta_2\Bigr) \\[6pt]
\lambda_3 &= 6\beta_2 -6\beta_1 +\beta_0 = \frac{\alpha}{\kappa}\Bigl(\beta_1 - 3\beta_2 + 2\beta_3\Bigr) \\
\lambda_4 &= 20\beta_3 -30\beta_2 +12\beta_1-\beta_0\\
&= \frac{\alpha}{\kappa}\Bigl(\beta_1 - 6\beta_2 + 12\beta_3 - 7\beta_4\Bigr).
\end{aligned}
\end{equation}

\section{Fresnel zone extraction}\label{sec-appen-F}

For a link with transmitter–obstacle distance $d_1$ and obstacle–receiver distance $d_2$, the radius of the
(n)-th Fresnel zone at the obstacle location is defined by 
 \begin{equation}
r_n = \sqrt{ \frac{n \lambda d_1 d_2}{d_1 + d_2} }.
\label{eq:fresnel_formula}
\end{equation}
To estimate the geographical size of the graphene-coated blocker stand capable of obstructing the LOS signal, a Fresnel zone-based simulation is performed. The transmission frequency is configured to $f=2.4$\,GHz, which is the typical carrier frequency of BLE devices. Given the speed of light, $C=3\times10^8$\,m/s, the corresponding wavelength can be calculated as $\lambda=c/f=12.5$\,cm, which is roughly $1/20$ smaller than the distance between a transmitter-receiver pair, confirming the planar wavefront assumption in the far-field. 
\begin{figure}[!ht]
\centering
\subfloat[Six-Fresnel-zone]{%
  \includegraphics[width=0.40\columnwidth]{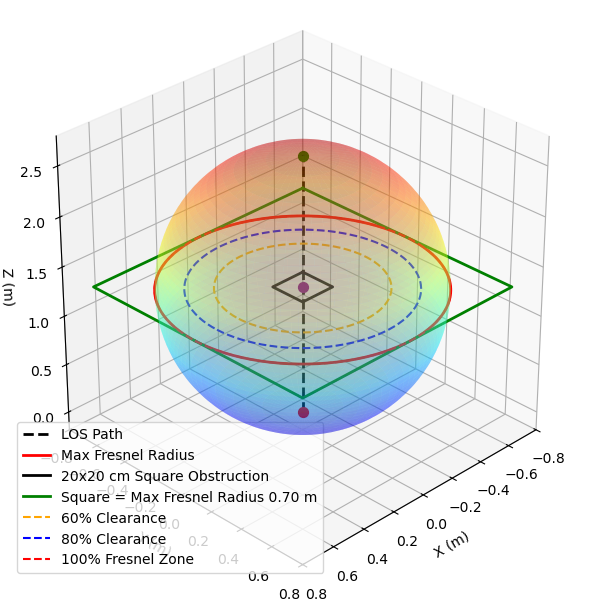}%
  \label{fig:sim_fresnel6}%
}\hfill
\subfloat[Graphe layer RSSI reduction]{%
  \includegraphics[width=0.55\columnwidth]{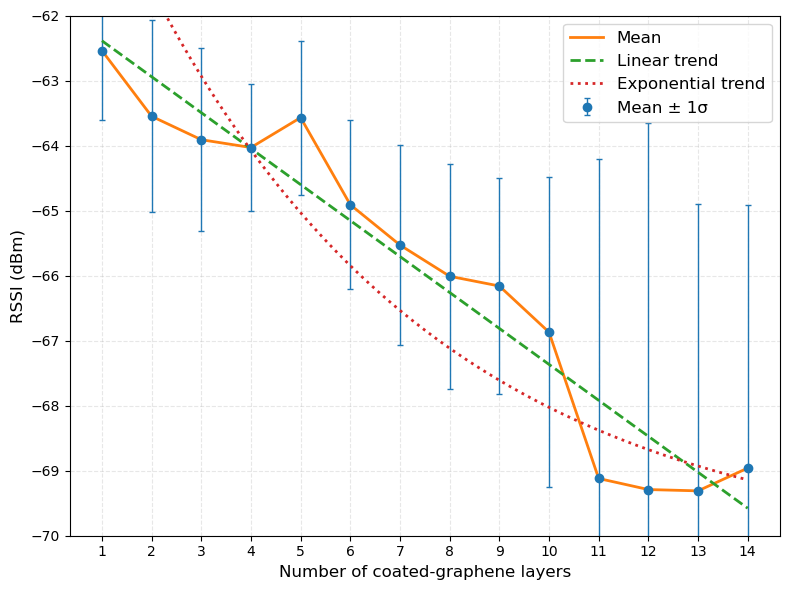}%
  \label{fig:rssi_layers}%
}\hfill
\caption{Graphene-induced NLOS emulation.}
\label{fig:exp_rssi}
\end{figure}
In Eq.~\eqref{eq:fresnel_formula}, a blocker of 20\,cm radius is sufficient for blocking the first Fresnel zone. Assuming that a blocker must intrude into at least more than two Fresnel zones, $n=6$, and the transmitter-receiver distance is $260$\,cm, a square object with dimensions approximately $70$\,cm is sufficient to block the LOS signal. To illustrate the attenuation (insulation) capability of the graphene-coated layers, an RSSI measurement campaign was conducted by progressively increasing the number of layers, while collecting 100 CTE packets in each measurement event, as depicted in Fig.~\ref{fig:exp_rssi}. The results show a clear reduction in the received RSSI from approximately -60\,dBm to -70\,dBm. A link-budget simulation combining the free-space path-loss model with a single-knife-edge diffraction obstacle loss \cite[Sec. 4.1]{ituR_P526_16_2025} yields an approximate received power of $\approx -72$\,dBm at the receiver (with 23.64\,dB diffraction loss), which is consistent with the measured RSSI levels. This agreement supports the physical interpretation that the graphene-coated blocker effectively suppresses the dominant LOS component and induces an NLOS propagation condition in which diffraction becomes the prevailing transmission mechanism. Additionally, the experiment is accompanied by increased variability at higher layer counts. This behavior is consistent with a multipath-dominated regime, where the superposition of reflected components yields alternating constructive and destructive interference, thereby amplifying RSSI fluctuations as the direct component is increasingly suppressed.

\bibliographystyle{IEEEtran}  
\bibliography{references}


 




\end{document}